\documentclass[twocolumn,preprintnumbers,prb,amssymb]{revtex4}
\usepackage{dcolumn}
\usepackage{bm}
\usepackage{verbatim}       

\usepackage[dvips]{graphicx}
\usepackage{amsthm}
\usepackage{amssymb}
\usepackage{bm}
\usepackage{amstext}
\usepackage{psfrag}
\usepackage{amsmath}
\usepackage{amsfonts}

\newcommand{\dd}{{\rm d}}

\begin{document}
%

\title{Universal one-way light speed from a universal light speed over closed paths}


\author{Ettore Minguzzi}
\affiliation{Dipartimento di Fisica dell'Universit\`a di Milano-Bicocca,\\ Piazza della Scienza 3, 20126 Milano, Italy \\ minguzzi@mib.infn.it}

\author{Alan Macdonald}
\affiliation{Department of Mathematics, Luther College, \\
Decorah, Iowa 52101 \\ macdonal@luther.edu}


%
%
%
%
%
%
%

\begin{abstract}
\noindent
 This paper gives two complete and elementary proofs that
if the speed of light over closed paths has a universal value $c$,
then it is possible to synchronize clocks in such a way that the
one-way speed of light is $c$. The first proof is an elementary
version of a recent proof. The second provides high precision
experimental evidence that it is possible to synchronize clocks in
such a way that the one-way speed of light has a universal value.
We also discuss an old incomplete proof by Weyl which is important
from an historical perspective.\\
\end{abstract}
\maketitle



\section{Introduction}
There has been much confusion about the relationship between
Einstein's definition of synchronized clocks \footnote{In this
paper we are interested only in the standard Einstein
synchronization. We do not enter into the debate over the
conventionality of synchronization \cite{anderson98,janis02}.} and
his postulate of the universality of the one-way speed of light.
Some authors \cite{resnik68, ellis00} assume a universal one-way
speed of light before discussing synchronization. We believe that
this is a logical error, as a one-way speed has no meaning until
clocks are synchronized. For the one-way speed of light from a
point of space $A$ to a point $B$ is defined as
$\overline{AB}/(t_{B}-t_{A})$, where $t_{A}$ is the time of
departure of a light beam from $A$ as measured by a clock at $A$,
and $t_{B}$ is the time of its arrival at $B$ as measured by the
clock at $B$. For this definition to be meaningful, the two clocks
need to be synchronized.

Other authors \cite{french68, Brehme88} state that once clocks
have been synchronized according to Einstein's definition,
then the speed of light is universal, that is,
independent of the point of space, of time, or of the direction
followed by a light beam.
However, this is not true, as the example of a Newtonian spacetime
with an ether frame shows \cite{macdonald83}.
Thus the assumption of a universal one-way light speed $c$ is actually two assumptions,
which together we call $\bf{1c}$:
\begin{itemize}
\item[(a)]  Clocks can be set so that every pair of them is Einstein synchronized.
\item[(b)]  The one-way speed of light with respect to the synchronized clocks
is a universal constant $c$.
\end{itemize}

Let us denote by $\bf{L/c}$ the assumption
of a universal light speed $c$ around closed paths.
Note that this is a synchronization independent concept
since for the measure of such an average speed only one clock is
required.

Our purpose here is to discuss this theorem, expressed in the title of this paper:
\smallskip

\textbf{Theorem.} $\bf{L/c} \Rightarrow \bf{1c}$.
\smallskip

We shall give two separate proofs of the theorem. The first is the
most direct. It is a considerable simplification of a proof of one
of the authors \cite{minguzzi02}. The second is based on work of
the other author \cite{macdonald83}. It provides high precision
experimental evidence for $\bf L/c$ and thus, by the theorem, for
$\bf 1c$. Finally, we discuss a proof given by Hermann Weyl in his
book ``Raum, Zeit, Materie" \cite{weylG}. We point out a tacit
assumption in the proof, which makes it incomplete, and then show
that the assumption follows from Weyl's other assumptions.

All this provides new logical and experimental insights into the
foundations of special relativity, since it justifies the fundamental
assumption of a universal one-way speed of light.

\section{The properties} \label{sec:properties}
We consider observers at rest with respect to each other, and
assume that their space is Euclidean. Spatial points are denoted
with letters $A,B,C \ldots$. Next, we assume that light propagates
on straight lines and that if a beam leaves a point $A$ at time
$t_{A}$, with respect to A's clock, directed toward $B$, it
reaches $B$ at a finite time $t_{B}$ with respect to $B$'s clock.

Einstein defined ``clocks at $A$ and $B$ are synchronized" as
follows. Emit a flash of light from $A$ to $B$ at time $t_{A}$.
Let it arrive at $B$ at time $t_{B}$.  Similarly, let a flash
emitted from $B$ at time $t_{B}'$ arrive at $A$ at time $t_{A}'$.
Say the clocks are {\em synchronized} if
\begin{equation} \label{synchronization}
t_{B}-t_{A}=t_{A}'-t_{B}'
\end{equation}
for all times $t_{A}$ and $t_{B}'$. If clocks can be set so that
this equation holds for any pair of them, then we say that
Einstein synchronization can be applied consistently.
When this is the case, we can define a global time
$t$: the time $t$ of an event at $P$ is the time of the event
according to the clock at $P$. We shall call this \emph{Einstein
time}.

Notice that with respect to Einstein time the one-way speed of
light between two points is equal to the two-way speed  between
the points. Indeed Eq. (\ref{synchronization}) states that the
time needed by light to go in direction $AB$ is the same as that
needed to go in the opposite direction $BA$.

In what follows we shall relate a number of properties which we
list here, adding mnemonics to the left.
\begin{itemize}
\item[{\bf z=0}.] Emit flashes of light from $B$ at times $t_{1B}$ and $t_{2B}$
according to a clock at $B$. Let them arrive at  $A$ at times
$t_{1A}$ and $t_{2A}$ according to a clock at $A$. Then
\begin{equation} \label{eq:z0}
t_{2A}- t_{1A}= t_{2B}-t_{1B}.
\end{equation}
This property is another way of saying that there is no redshift.
\item[$\pmb{\triangle}$.] The time it takes light to traverse a
triangle (through reflections over suitable mirrors) is
independent of the direction taken around the triangle.
\item[$\bf{2c}$.] The two-way speed of light has a constant value $c$.
\item[$\bf{L/c}$.] The time it takes light to traverse a closed polygonal path
(through reflections over suitable mirrors) of length $L$ is
$L/c$, where $c$ is a constant.
\item[\bf{syn}.] Einstein synchronization can be applied consistently.
\item[$\bf{1c}$.] Einstein synchronization can be applied consistently
and the one-way speed of light with respect to the synchronized clocks
has a constant value $c$.
\end{itemize}

A constant $c$ appears in the definitions of $\bf{2c}$, $\bf{L/c}$,
and $\bf{1c}$. As we shall see in the proofs below, the
implications among these properties refer to the same value of
$c$. This will justify our notation.

Notice that the first four properties do not depend on clock synchronization.

Figure 1 summarizes the implications we will establish between these properties.

\begin{figure}[ht] \label{fig:implications}
\centering \psfrag{1c}{\Large{$1c$}} \psfrag{2c}{\Large{$2c$}}
\psfrag{L/c}{\Large{$L/c$}} \psfrag{z=0}{\Large{$z=0$}}
\psfrag{A}{\Large{$\triangle$}} \psfrag{syn}{\Large{syn}}
\includegraphics[width=5cm]{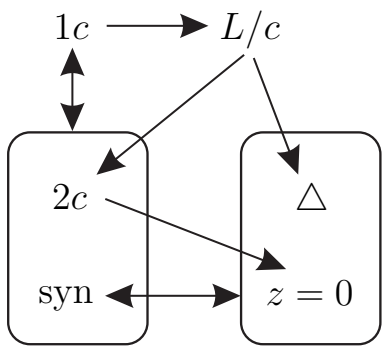}
\caption{Implications proved in the text.}
\end{figure}

We can immediately establish several trivial implications:

$\bf{1c} \Rightarrow \bf{L/c}$. This requires no comment.

$\bf{L/c} \Rightarrow \bf{2c}$. A light beam that goes from $A$ to
$B$ and back traverses a closed path. According to $\bf{L/c}$, the
two way speed is $c$.

$\bf{L/c} \Rightarrow \pmb{\triangle}$. According to
$\bf{L/c}$, the time to traverse a triangle is $L/c$, which is
independent of the direction taken around the triangle.

$\{ \bf{syn} \ \textrm{and} \ \bf{2c} \} \Leftrightarrow \bf{1c}$.
We have  noted that if {\bf syn} holds,
then the two-way speed is equal to the one-way speed.
Since from $\bf{2c}$ the two-way speed is $c$, the one-way speed is $c$.
The equivalence now follows easily.

The only implications remaining are
${\bf 2c} \Rightarrow ({\bf z=0})$ and
$\textrm{\bf{syn}} \Leftrightarrow \{ ({\bf z=0}) \mathrm{\ and\ } \pmb{\triangle}\}$.
They will be established as part of the proof of
$\bf{L/c} \Rightarrow \bf{1c}$ in Section \ref{sec:proof2}.

\section{The most direct proof} \label{sec:proof1}
A proof of the theorem $\bf{L/c} \Rightarrow \bf{1c}$ was
given by one of the authors in \cite{minguzzi02}.
The elementary nature of the proof was
however hidden by an infinitesimal approach. The strategy was to
 assign a label, an ``Einstein time" to any event and, in the
end, to synchronize clocks. We give here an elementary version of
that proof.

Consider a point $O$ and a clock
at rest at $O$. We define a time $t(e)$ of a generic event $e$.
Emit a light beam from $O$ to $e$, where it is reflected back to $O$.
\footnote{ We show that given
two points $A$ and $B$, and a time $t_{1B}$, a light beam can be
sent from $A$ that reaches $B$  at $t_{1B}$. Define $t_{0B}$ by
$t_{1B} - t_{0B} = 2\overline{AB}/c$. Emit a light beam from $B$
at time $t_{0B}$ toward $A$. At $A$ reflect it back to $B$.  Then
from $\bf{L/c}$, after the reflection at $A$ the beam  returns to
$B$ at $t_{1B}$.}
Let the departure and arrival times of the
light beam be $t^{i}(e)$ and $t^{f}(e)$ according to the clock at
$O$. Define
\begin{equation}
t(e)= \frac{t^{i}(e)+t^{f}(e)}{2}.
\end{equation}
This procedure assigns a time $t(e)$ to every event $e$.
The time $t$ at $O$ is measured by the clock at $O$.
The procedure also defines
times $t^i(e)$ and $t^f(e)$ for every event $e$. From $\bf{L/c}$,
if $e$ is at point $E$, then $t^{f}(e)-t^{i}(e) =
2\overline{OE}/c$.

Let us prove that the one way speed of light with respect to $t$
is $c$. Emit a light beam from point A at event $a$. Let it arrive
at point $B$ at event $b$. We can imagine that the beam starts at
$O$, arrives at $a$, is reflected to $b$, and is reflected back to
$O$. From $\bf{L/c}$ we have
\begin{eqnarray}
t^{f}(b)-t^{i}(a)&=&\overline{OABO}/c\, , \\
t^{f}(a)-t^{i}(a)&=&2\overline{OA}/c\, , \\
t^{f}(b)-t^{i}(b)&=&2\overline{OB}/c\, .
\end{eqnarray}
Subtracting the second and third equations from twice the first
gives
\begin{eqnarray}
t(b)-t(a)=\overline{AB}/c \, ;
\end{eqnarray}
the one way speed of light with respect to $t$ is $c$.

Now consider another ``global time" $\tilde{t}$ established by a
clock at point $\tilde{O}$. Since light moves at speed $c$ with
respect to both $t$ and $\tilde{t}$, one has that if $a$ and $b$
are events on the world line of a light beam, then
\begin{equation}
t(b)-\tilde{t}(b)=t(a)-\tilde{t}(a).
\end{equation}
Since light moves at constant speed with respect to $t$, given any
two events whatsoever $e$ and $f$, there is a third event $g$ in
the intersection of their light cones. Therefore
\begin{equation}
t(e)-\tilde{t}(e)=t(g)-\tilde{t}(g)=t(f)-\tilde{t}(f).
\end{equation}
Now if we fix $f$ and let $\emph{e}$ vary, then we see that
$t(e)=\tilde{t}(e)+const$. Thus we can take $\tilde{t} = t$ by
resetting the clock at $\tilde{O}$ which defines $\tilde{t}$. The
definition of global time is therefore independent of the initial
point $O$ chosen.

The time $\tilde{t}$ at $\tilde{O}$ is measured by the clock at $\tilde{O}$.
Since $\tilde{t} = t$ at $\tilde{O}$,
the time $t$ at $\tilde{O}$ is also measured by the clock at $\tilde{O}$.

Thus $\bf{L/c}$ implies that clocks can be synchronized using
Einstein's method without bothering about an origin $O$. Moreover
the one way speed of light with respect to the synchronized clocks is $c$.

\section{A proof providing experimental evidence for $\bf{L/c}$}
\label{sec:proof2} The proof in this section has two advantages.
First, it passes through {\bf syn}, thus making clear its relation
to ${\bf L/c}$. Second, it provides high precision experimental
evidence for ${\bf L/c}$.

We noted at the end of Section \ref{sec:properties}
that of the implications in Figure 1, only
$\textrm{\bf{syn}} \Leftrightarrow \{ (\,{\bf z=0}) \mathrm{\ and\ } \pmb{\triangle}\}$
and ${\bf 2c} \Rightarrow (\,{\bf z=0})$
remain to be established. We shall do this momentarily.
Then following the arrows in Figure 1 gives our second proof of
${\bf L/c} \Rightarrow {\bf 1c}$.

$\textrm{\bf{syn}} \Leftrightarrow \{ (\,{\bf z=0}) \mathrm{\ and\ } \pmb{\triangle}\}$.
This characterization of ${\bf syn}$ has been proved by one of the authors in \cite{macdonald83}.

${\bf 2c} \Rightarrow (\,{\bf z=0})$.
This result appears to be new.
Consider two points $A$ and $B$ and
arbitrary times $t_{1B}$ and $t_{2B} > t_{1B}$ according to a clock at $B$.
We must show that if light beams are sent from $B$ to $A$ at times
$t_{1B}$ and $t_{2B}$, arriving at $A$ at times
$t_{1A}$ and $t_{2A}$  according to a clock at $A$, then Eq. (\ref{eq:z0}) holds.

\begin{figure}[ht] \label{fig:2cz0base}
\centering \psfrag{2A}{\Large{$t_{2A}$}}
\psfrag{1A}{\Large{$t_{1A}$}} \psfrag{0A}{\Large{$t_{0A}$}}
\psfrag{2B}{\Large{$t_{2B}$}} \psfrag{1B}{\Large{$t_{1B}$}}
\psfrag{0B}{\Large{$t_{0B}$}} \psfrag{A}{\Large{$A$}}
\psfrag{B}{\Large{$B$}} \psfrag{C}{\Large{$C$}}
\includegraphics[width=4cm]{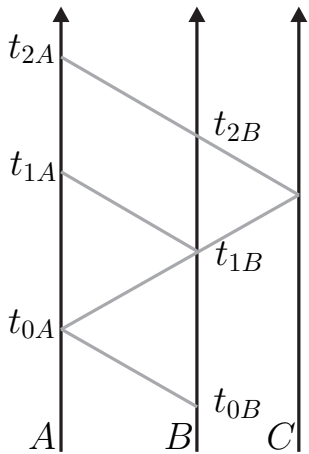}
\caption{The reflections considered in $\bf{2c} \Rightarrow
(\bf{z=0})$.}
\end{figure}

Refer to Figure 2. Emit a light beam from $B$ at time $t_{0B}$
toward $A$. At $A$ reflect it back to $B$. Define $t_{0B}$ by
$t_{1B} - t_{0B} = 2\overline{AB}/c$. Then from $\bf 2c$, the beam
will return to $B$ at $t_{1B}$. At $B$ the beam is split by a
semi-transparent mirror. The reflected part of the beam arrives
back at $A$ at time $t_{1A}$. The transmitted part arrives at a
point $C$, where it is reflected back to $B$. Choose $C$ so that
$\overline{BC}$ by $t_{2B} - t_{1B} = 2\overline{BC}/c$. Then from
$\bf 2c$ the beam will return to $B$ at $t_{2B}$. At $B$ the beam
again encounters the semi-transparent mirror. The transmitted part
arrives at $A$ at $t_{2A}$. We consider the  paths $ABA$, $ACA$,
and $BCB$. Since $\bf{2c}$ holds:
\begin{eqnarray}
ABA &\to& t_{1A}-t_{0A}= 2 \overline{AB}/c, \\
ACA &\to& t_{2A}-t_{0A}= 2 \overline{AC}/c, \\
BCB &\to& t_{2B}-t_{1B}= 2 \overline{BC}/c.
\end{eqnarray}
Summing the first and third equations and subtracting the second gives Eq. (\ref{eq:z0}).
This completes our proof.

\subsection{Experimental evidence}
Following the implications in
Figure 1, we see that
\begin{equation}
\bf{L/c} \Leftrightarrow \{ \bf{2c} \ \textrm{and} \ \pmb{\triangle} \}
         \Leftrightarrow \bf{1c}\,.
\end{equation}
The implications ultimately rest upon our result $\bf{2c} \Rightarrow (\bf{\,z=0})$\,.
They show that evidence for $\bf{2c}$ or $\pmb{\triangle}$
is evidence for $\bf{L/c}$ and $\bf{1c}$.
A direct test of \textbf{L/c} requires a ruler-measurement of $L$ for several paths.
We know of no high precision test.
On the other hand, tests of $\bf{2c}$ and $\pmb{\triangle}$,
which we are about to discuss, are interferometric, and thus of high precision.

$\bf{2c}.$ There are well known tests of $\bf{2c}$. The
Michelson-Morley experiment shows that the two-way speed of light
is the same in perpendicular directions from a point. This result
is consistent with $\bf{2c}$. But it is not conclusive: the
equal-in-perpendicular-directions speed could still be different
at different times, places, or speeds, violating $\bf{2c}$. The
Kennedy-Thorndike experiment was performed to eliminate this
possibility. Together the experiments test $\bf{2c}$. A modern
version of both experiments with an impressive accuracy has
recently been performed \cite{wolf03}.

$\pmb{\triangle}$. We know of no direct test of $\pmb{\triangle}$.
But an experiment of Macek and Davis using a ring laser in the
shape of a square shows that the time it takes light to traverse
the square is independent of the direction taken around it to one
part in $10^{12}$ \cite{macek63}. There is no reason that the
experiment could not be performed with a ring laser in the shape
of a triangle. For further discussion, see \cite{erlichson73,
stedman73}.

The ring laser experiment support for $\pmb{\triangle}$
is consistent with $\bf{L/c}$.
But it is not conclusive: our counterexamples in the appendix
show that $\pmb{\triangle}$ does not imply $\bf{L/c}$.
The equal-in-opposite-directions speed from $\pmb{\triangle}$
could still be different at different times, places, and speeds,
violating $\bf{L/c}$.
Our work eliminates this possibility.
It shows that $\pmb{\triangle}$ and $\bf{2c}$
together imply $\bf{L/c}$ to high precision.

We note that while $\bf{2c}$ is a consequence of $\bf{L/c}$,
our counterexamples show that $\bf{2c}$ does not imply $\bf{L/c}$.
Without evidence, one should be hesitant to extrapolate from $\bf{2c}$ to $\bf{L/c}$,
as light traversing paths which enclose an area
behaves differently from light traversing paths which do not.
An example is provided by the Michelson-Morley and Macek-Davis experiments.
If the apparatus of the Michelson-Morley experiment is rotating,
the interference fringes do not shift.
Indeed, the null result of the experiment is just this fact.
If the ring laser in the Macek-Davis experiment is rotating,
then the interference fringes do shift.
Indeed, the experiment was performed to test just this effect.

\section{Weyl's incomplete proof} \label{sec:weyl}
The fifth edition of ``Raum, Zeit, Materie" by Hermann Weyl contains,
among the other modifications, a new Section 23 devoted to an analysis
of the postulates of special relativity.
Here Weyl gives a proof of the theorem $\bf{L/c} \Rightarrow \bf{1c}$.

To the best of our knowledge, Weyl's proof has not been published
in English. (Henry Brose's English translation \cite{weylE} is of
the fourth edition.)
We thus provide a translation:
\begin{quote}
Section 23. Analysis of the Relativity Principle. The division of
the world into space and time as projection.\\

We want to as carefully as possible determine upon which
conditions and facts the validity of Einstein's relativity is
based. The Michelson experiment shows that in any uniformly moving
reference frame $K$ (which we imagine as a flat structure) the
proportionality Equation (20) [\,our ${\bf 2c}$] is valid with a
constant $c$. If $A$ and $B$ are two points in $K$, then a clock
at $B$ runs synchronized with a clock at $A$, whenever a light
signal, which at any given time $t$ is emitted at $A$, arrives at
$B$ at time $t + \overline{AB}/c$; that is the definition of
``synchronized". Conversely, as a matter of fact, the clock at $A$
is synchronized with $B$ (or, in short,  $A$ runs in the same way
as $B$). Moreover, we need to consider another circumstance, that
if $B$ runs as $A$ and $C$ as $B$, then $C$ runs as $A$; what does
that mean? Let us take two clocks set up at $A$; we synchronize
the clock at $B$, by means of light signals, with the first clock
at $A$; we synchronize the clock at $C$ with the clock at $B$; and
the second clock at $A$ with the clock at $C$; then our claim is
that the two clocks at $A$ constantly show the same time. The
second clock at $A$ shows the time $t + L/c$ when a light signal,
which is emitted from $A$ at the time $t$ at $A$ of the first
clock at $A$ and has traversed the path $ABCA$ of length $L$,
arrives back at $A$. Our claim thus says that the time which
passes between the departure and arrival of the light signal
equals $L/c$. It is reasonable to establish this fact not only for
the ``two sided triangle'' $ABA$, and the triangle $ABCA$, but
also for any polygonal path. We express it here as a first
experimental fact:

{\em If one lets the light, in a reference frame $K$ in uniform
translation, traverse a closed polygonal path of length $L$, then
between the departure and arrival of the light signal a time
$\tau$ elapses which is proportional to $L$: $\tau = L/c$. }

Accordingly, it is possible to introduce a time $t = x_0$ anywhere
in the reference frame; the clocks which show it all run
synchronously. They can be synchronized by means of light signals
from a central clock; this synchronizing is independent of the
center that is selected. We expressly note that in these cases
sources of light are always to be used which are at rest in $K$.
\end{quote}

Weyl starts his analysis from the experimental evidence for
$\bf{2c}$ and uses the constant $c$ to give a definition of
synchronization that differs from Einstein's definition. According
to Weyl's definition, a clock at $B$ is synchronized with a clock
at $A$ if a light beam that leaves $A$ at time $t$ with respect to
$A$'s clock reaches $B$ at time $t + \overline{AB}/c$ with respect
to $B$'s clock.
From $\bf{2c}$, this definition of synchronization is symmetric.
Note that with $\bf{2c}$ Einstein's and Weyl's definition are equivalent.

Weyl next proves the transitivity of synchronization. As his proof
is perhaps not completely transparent, we offer this explication.
Suppose that $B$'s clock is synchronized with $A$'s and $C$'s
clock is synchronized with $B$'s. We need to prove that $A$'s
clock is synchronized with $C$'s. Synchronize a second clock at
$A$ with $C$'s clock. Consider a light beam that leaves $A$ at
time $t_A$ according to $A$'s first clock and traverses the path
$ABCA$, returning to $A$ at time $t^r_{A'}$ according to $A$'s
second clock. From the given synchronizations,
\begin{equation}
t^r_{A'}-t_A=L/c,
\end{equation}
where $L$ is the length of the path $ABCA$.
Let $t^r_A$ be the time of the return of the light beam at $A$
according to $A$'s first clock.
Then from $L/c$
\begin{equation}
t^r_A-t_A=L/c.
\end{equation}
Thus
\begin{equation}
t^r_A=t^r_{A'},
\end{equation}
that is, the two clocks at $A$ are synchronized
(in the end they are the same clock),
and therefore $A$'s first clock is synchronized with $C$'s clock.

Clocks over space can be synchronized with respect to a central clock.
Since synchronization is transitive, the result does not depend
on the central clock chosen (up to a global resetting of clocks).\\

In order to be meaningful, Weyl's definition of synchronization
must be independent of the time the light beam is emitted,
i.e., the property $\bf z=0$ must hold.
Weyl apparently assumes this tacitly.
We have shown above that ${\bf 2c} \Rightarrow (\,{\bf z = 0}\!)$,
a fact of which Weyl was probably not aware.

Weyl, in the translation above, takes $\bf{L/c}$ as an ``experimental fact''.
However, as noted in Section \ref{sec:proof2},
we are unaware of any precision test of $\bf{L/c}$.

C. M{\o}ller has given a proof of $\bf 1c$ from two assumptions
\cite{Moller}. He proves his Eq. (2.1), which is $\bf 1c$. The two
assumptions are:

${\bf L/c}$. This is M{\o}ller's Eq. (2.3).
He justifies it by appeal to Fizeau's experiment.
But this experiment is a test of $\pmb{\triangle}$, not ${\bf L/c}$,
as we argued with respect to the ring laser experiment
at the end of Section \ref{sec:proof2}.

$\bf z=0$. This is M{\o}ller's condition 1.
Unlike Weyl, M{\o}ller realizes the relevance of $\bf z=0$.
He justifies it with these words:

\begin{quote}
[Condition 1] is no doubt fulfilled,
since all points in an inertial system are equivalent,
so that two standard clocks which have the same rate when placed together at $O$
will also have the same rate when they are installed at different points $O$ and $P$.
\end{quote}
The phrase ``have the same rate when they are installed at
different points'' expresses $\bf z=0$. His justification is spatial homogeneity:
``all points in an inertial system are equivalent''.
Thus according to M{\o}ller, $z=0$ in any spatially homogeneous space.
The following counterexample shows that this is wrong.
Consider a Newtonian spacetime with an absolute space and an absolute time $t$.
Suppose that the speed of light in the spacetime depends on $t:\ c = c(t)$.
This spacetime is spatially homogeneous and $z \ne 0$.
It thus appears to us that M{\o}ller provides no valid justification for $\bf z=0$.
We obtain $\bf z=0$ from ${\bf 2c} \Rightarrow (\,{\bf z = 0})$.

We conclude that we provide better evidence for $\bf{L/c}$
and a more complete proof of $\bf{L/c} \Rightarrow \bf{1c}$
than do Weyl or M{\o}ller.

\section*{Acknowledgements}
We thank Professor James Skurdall and Dr. Daniela Del Tessa  for
their help with the translation from Weyl's German text. Of course
any errors in the translation are our responsibility.

\section*{Appendix: counterexamples}

\renewcommand{\theequation}{A\arabic{equation}}
\setcounter{equation}{0}  

One can wonder whether there are, between the properties in
Figure 1, relations that are not obtainable from the figure
by simply following the arrows. The answer is
negative since the counterexamples, listed below,  show that
$\bf{2c} \nRightarrow \pmb{\triangle}$,
$({\bf z=0}) \nRightarrow \pmb{\triangle}$,
$\bf{syn} \nRightarrow \bf{2c}$,  and
$\pmb{\triangle} \nRightarrow ({\bf z=0})$.
It can be seen from the counterexamples and the figure
that there are no other implications among individual properties.

$\bf{2c} \nRightarrow \pmb{\triangle}$.
Consider a Newtonian spacetime with an absolute space and time.
Suppose that the velocity of light at the point $\bf x$
in the direction $\bf \hat{v}$ is
\begin{equation} \label{eq:velocity}
\frac{c \bf{\hat{v}}}{1 + c {\bf\hat{v}} \cdot {\bf A}({\bf x})}\ ,
\end{equation}
where ${\bf A}({\bf x})$ is a field such that
$\nabla \times {\bf A} \ne 0$ and $|{\bf A}|<1/\,2c$.
Then the time it takes light to traverse a closed path $\gamma$
of length $L$ is \cite{minguzzi02}
\begin{equation} \label{eq:rp}
\Delta t=\frac{L}{c} + \,\oint_{\gamma} {\bf A} \cdot \dd {\bf l}\, .
\end{equation}

Given $\bf x$, let $\gamma$ be an equilateral triangle centered at $\bf x$
and orthogonal to $\nabla \times {\bf A}({\bf x})$.
If $\gamma$ is small enough, then the integral in Eq. (\ref{eq:rp}) is not zero.
Thus the round trip time $\Delta t$ depends on the direction
followed by the light, and hence $\pmb{\triangle}$ does not hold.
However, $\bf{2c}$ holds because the integral vanishes for the
path that goes from $A$ to $B$ and back.

$({\bf z=0}) \nRightarrow \pmb{\triangle}$. Consider again the
setting of the last counterexample. We have already noticed that
with the speed Eq. (\ref{eq:velocity}), $\pmb{\triangle}$ does not hold.
Moreover, the speed of light at a given point does not change in time.
Thus the worldlines of light that goes from
$A$ to $B$ are time translations of each other, and so $z=0$.

$\bf{syn} \nRightarrow \bf{2c}$. Notice that the definition of
$\bf{syn}$  does not involve any metric over the space. Start from
Minkowski spacetime. Taking unaltered its light cone structure in
the coordinates $t, x, y ,z$ redefine the the space metric to be
$\dd l^{2} = k^{2} \dd x^{2}+ \dd y^{2}+\dd z^{2}$. You have
obtained a space where $\bf{syn}$ still holds but the two-way
speed of light is anisotropic since that in the $x$ direction is
$k c$ whereas that in the $y$ direction is $c$.

Another counterexample \cite{macdonald83} comes from the old ether
theory of the propagation of light. In such a theory the light
propagates at a constant speed with respect to the ether. Inertial
observers in motion with respect to the ether can apply
consistently the Einstein synchronization method but the one-way
speed turns out to be anisotropic since the two-way speed itself,
in those frames,  is anisotropic.

$\pmb{\triangle} \nRightarrow ({\bf z=0})$. Consider Minkowski
spacetime with the usual coordinates $\{x^{i}, t\}$. Suppose,
however, that the clocks at rest do not measure $t$ but $t'=t
(1+r^{2}/a^{2})$ where $r$ is the radial distance from the origin,
and $a$ is a constant. In this space  $\pmb{\triangle}$
still holds since it involves only one clock.
On the contrary, $z \ne 0$.


\begin{thebibliography}{10}
\expandafter\ifx\csname
natexlab\endcsname\relax\def\natexlab#1{#1}\fi
\expandafter\ifx\csname bibnamefont\endcsname\relax
  \def\bibnamefont#1{#1}\fi
\expandafter\ifx\csname bibfnamefont\endcsname\relax
  \def\bibfnamefont#1{#1}\fi
\expandafter\ifx\csname citenamefont\endcsname\relax
  \def\citenamefont#1{#1}\fi
\expandafter\ifx\csname url\endcsname\relax
  \def\url#1{\texttt{#1}}\fi
\expandafter\ifx\csname
urlprefix\endcsname\relax\def\urlprefix{URL }\fi
\providecommand{\bibinfo}[2]{#2}
\providecommand{\eprint}[2][]{\url{#2}}

\bibitem{resnik68}
\bibinfo{author}{\bibfnamefont{R.}~\bibnamefont{Resnik}},
  \emph{\bibinfo{title}{Introduction To Special Relativity}}
  (\bibinfo{publisher}{John {W}iley {\&} Sons}, \bibinfo{address}{New York},
  \bibinfo{year}{1968}).

\bibitem{ellis00}
\bibinfo{author}{\bibfnamefont{G.~F.~R.} \bibnamefont{Ellis}} \bibnamefont{and}
  \bibinfo{author}{\bibfnamefont{R.~M.} \bibnamefont{Williams}},
  \emph{\bibinfo{title}{Flat and curved space-times}}
  (\bibinfo{publisher}{Oxford {U}niversity {P}ress}, \bibinfo{address}{Oxford},
  \bibinfo{year}{2000}), \bibinfo{edition}{2nd} ed.

\bibitem{french68}
\bibinfo{author}{\bibfnamefont{A.~P.} \bibnamefont{French}},
  \emph{\bibinfo{title}{Special Relativity}} (\bibinfo{publisher}{Chapman and
  Hall}, \bibinfo{address}{London}, \bibinfo{year}{1968}).

\bibitem{Brehme88}
\bibinfo{author}{\bibfnamefont{R.~W.} \bibnamefont{Brehme}},
  \emph{\bibinfo{journal}{Am. J. Phys.}} \textbf{\bibinfo{volume}{56}},
  \bibinfo{pages}{811} (\bibinfo{year}{1988}).

\bibitem{macdonald83}
\bibinfo{author}{\bibfnamefont{A.~L.} \bibnamefont{Macdonald}},
  \emph{\bibinfo{journal}{Am. J. Phys.}} \textbf{\bibinfo{volume}{51}},
  \bibinfo{pages}{795} (\bibinfo{year}{1983}).

\bibitem{minguzzi02}
\bibinfo{author}{\bibfnamefont{E.}~\bibnamefont{Minguzzi}},
  \emph{\bibinfo{journal}{Found. Phys. Lett.}} \textbf{\bibinfo{volume}{15}},
  \bibinfo{pages}{153} (\bibinfo{year}{2002}).

\bibitem{weylG}
\bibinfo{author}{\bibfnamefont{H.}~\bibnamefont{Weyl}},
  \emph{\bibinfo{title}{Raum, Zeit, Materie}}
  (\bibinfo{publisher}{{Springer-Verlag}}, \bibinfo{address}{New York},
  \bibinfo{year}{1988}),
  \bibinfo{note}{seventh edition based on the fifth {G}erman edition (1923)}.

\bibitem{anderson98}
  \bibinfo{author}{\bibfnamefont{R.}~\bibnamefont{Anderson}},
  \bibinfo{author}{\bibfnamefont{I.}~\bibnamefont{Vetharaniam}},
       \bibnamefont{and}
  \bibinfo{author}{\bibfnamefont{G.~E.}~\bibnamefont{Stedman}},
  \emph{\bibinfo{journal}{Phys. Rep.}} \textbf{\bibinfo{volume}{295}},
  \bibinfo{pages}{93} (\bibinfo{year}{1998}).

\bibitem{janis02}
\bibinfo{author}{\bibfnamefont{A.} \bibnamefont{Janis}},
 \bibinfo{title}{Conventionality of Simultaneity},
  \emph{\bibinfo{journal}{The Stanford Encyclopedia of Philosophy}},
  \bibinfo{edition}{(Fall 2002 Edition) E. N. Zalta (ed.)},
  \bibinfo{note}{http://plato.stanford.edu/archives/fall2002/entries/spacetime-convensimul/}.

\bibitem{wolf03}
  \bibinfo{author}{\bibfnamefont{P.}~\bibnamefont{Wolf}},
  \bibinfo{author}{\bibfnamefont{S.}~\bibnamefont{Bize}},
  \bibinfo{author}{\bibfnamefont{A.}~\bibnamefont{Clairon}},
  \bibinfo{author}{\bibfnamefont{A.~N.} \bibnamefont{Luiten}},
  \bibinfo{author}{\bibfnamefont{G.}~\bibnamefont{Santarelli}},
  \bibnamefont{and} \bibinfo{author}{\bibfnamefont{M.~E.} \bibnamefont{Tobar}},
  \emph{\bibinfo{journal}{Phys. Rev. Lett.}} \textbf{\bibinfo{volume}{90}},
  \bibinfo{pages}{060402} (\bibinfo{year}{2003}).

\bibitem{macek63}
\bibinfo{author}{\bibfnamefont{W.~M.} \bibnamefont{Macek}} \bibnamefont{and}
  \bibinfo{author}{\bibfnamefont{D.~T.~M.} \bibnamefont{{Davis, Jr.}}},
  \emph{\bibinfo{journal}{Appl. Phys. Lett.}} \textbf{\bibinfo{volume}{2}},
  \bibinfo{pages}{67} (\bibinfo{year}{1963}).

\bibitem{erlichson73}
\bibinfo{author}{\bibfnamefont{H.} \bibnamefont{Erlichson}},
  \emph{\bibinfo{journal}{Am. J. Phys.}} \textbf{\bibinfo{volume}{41}},
  \bibinfo{pages}{1298} (\bibinfo{year}{1973}).

\bibitem{stedman73}
\bibinfo{author}{\bibfnamefont{G.~E.} \bibnamefont{Stedman}},
  \emph{\bibinfo{journal}{Am. J. Phys.}} \textbf{\bibinfo{volume}{41}},
  \bibinfo{pages}{1300} (\bibinfo{year}{1973}).

\bibitem{weylE}
\bibinfo{author}{\bibfnamefont{H.}~\bibnamefont{Weyl}},
  \emph{\bibinfo{title}{Space Time Matter}} (\bibinfo{publisher}{Dover
  {P}ublications, {I}nc.}, \bibinfo{address}{New York}, \bibinfo{year}{1952}),
  \bibinfo{note}{based on the fourth {G}erman edition (1921) translation by
  {H}. {L}. {B}rose}.

\bibitem{Moller}
\bibinfo{author}{\bibfnamefont{C.}~\bibnamefont{M{\o}ller}},
  \emph{\bibinfo{title}{The Theory of Relativity, }}
  (\bibinfo{publisher}{{Clarendon Press}}, \bibinfo{address}{Oxford},
  \bibinfo{year}{1972}), \bibinfo{edition}{2nd} ed, Sec. 2.2.

\end{thebibliography}
\end{document}